# Imaging stacking order in few-layer graphene

Chun Hung Lui[1], Zhiqiang Li[1], Zheyuan Chen[2], Paul V. Klimov[1,†], Louis E. Brus[2], and Tony F. Heinz[1,*]

[1] *Departments of Physics and Electrical Engineering, Columbia University, 538 West 120th Street, New York, New York 10027, USA.*

[2] *Department of Chemistry, Columbia University, 3000 Broadway, New York, New York 10027, USA*

* *To whom correspondence should be addressed. E-mail: tony.heinz@columbia.edu*

† *Current address: Center for Spintronics and Quantum Computation, University of California, Santa Barbara, California 93106, USA*

**ABSTRACT** Few-layer graphene (FLG) has been predicted to exist in various crystallographic stacking sequences, which can strongly influence their electronic properties. We demonstrate an accurate and efficient method to characterize stacking order in FLG using the distinctive features of the Raman 2D-mode. Raman imaging allows us to visualize directly the spatial distribution of Bernal (ABA) and rhombohedral (ABC) stacking in tri- and tetra-layer graphene. We find that 15% of exfoliated graphene tri- and tetra-layers is comprised of micron-sized domains of rhombohedral stacking, rather than of usual Bernal stacking. These domains are stable and remain unchanged for temperatures exceeding 800 ºC.

Graphene-based materials have stimulated intense interest because of their remarkable electronic properties and potential for novel applications. With the impressive progress in research on graphene mono- and bi-layers, recent attention has also turned to graphene's few-layer counterparts[1-7]. In few-layer graphene (FLG), the crystallographic stacking of the individual graphene sheets provides an additional degree of freedom[4, 8-11]. The distinct lattice symmetries associated with different stacking orders of FLG have been predicted to strongly influence the electronic properties of FLG[3, 4, 8-23], including the band structure[8, 10, 12-19], interlayer screening[20], magnetic state[22, 23], and spin-orbit coupling[21]. Experimentally, the strong influence of stacking order on the low-energy electronic structure of FLG was recently demonstrated by infrared (IR) spectroscopy[4]. For graphene trilayers, two stable crystallographic configurations are predicted: ABA and ABC stacking order[8-10, 12, 13] (Figure 1). In the absence of direct evidence of ABC stacking order in trilayers, ABA stacking order has generally been presumed in most studies of exfoliated materials, as this structure is believed to be slightly more stable thermodymically than the ABC stacking order. Recent studies[3, 8-10, 12, 14, 16, 17, 19, 20, 24], indicate, however, distinct properties for these two types of graphene trilayers. ABA-stacked trilayers are semi-metals with an electrically tunable band overlap[3, 10, 12, 16-18], while ABC-stacked trilayers are predicted to be semiconductors with an electrically tunable band gap[10, 12, 15, 19]. In view of these differences, research on FLG requires the development of convenient and accurate methods for characterizing stacking order and its spatial distribution.

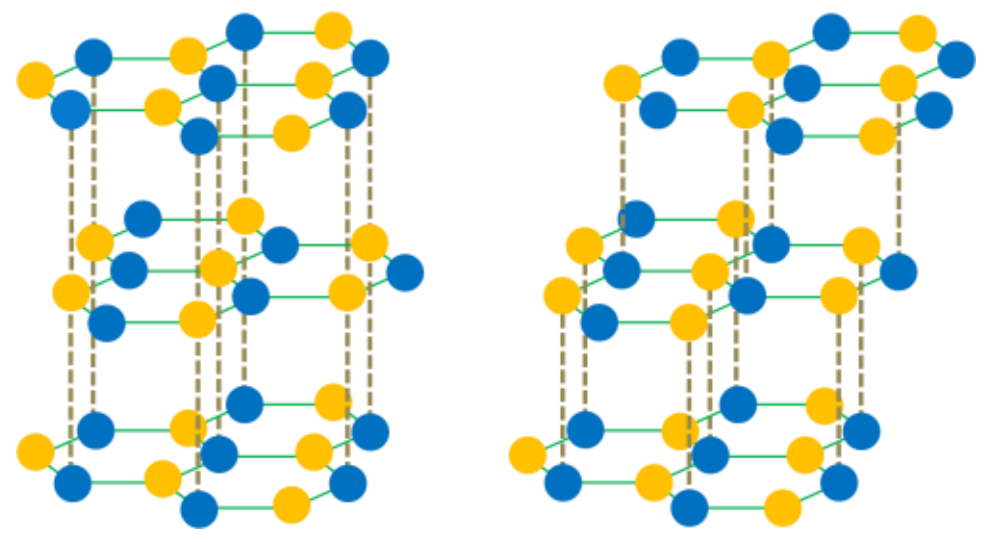

**Figure 1.** Lattice structure of trilayer graphene with ABA (left) and ABC (right) stacking sequence. The yellow and blue dots represent carbon atoms in the A and B sub-lattices of the graphene honeycomb structure.

While IR spectroscopy provides a means of identifying stacking order in FLG[4], it requires somewhat specialized instrumentation and cannot provide high spatial resolution. Raman spectroscopy, on the other hand, has the potential to overcome these limitations and serves as an effective general approach for characterization and spatial imaging of stacking order. The technique has already proved to be a reliable and efficient method for determining many physical properties of graphene layers[6, 25]. The intensity of D-mode indicates the defect density[25]; the peak position and lineshape of G-mode reflect the doping[25, 26] and strain level[27, 28]. In addition, the 2D (G') mode, arising from a double-resonant electronic process[6, 25, 29, 30], is sensitive not only to the vibrational features of graphene, but also to its electronic structure. As such, its lineshape provides an accurate signature of graphene mono- and bi-layers[6, 25, 31, 32].

In this Letter, we demonstrate that stacking order in tri- and tetra-layer graphene samples can be readily identified by means of Raman spectroscopy. We find that both Bernal (ABA) and rhombohedral (ABC) stacking order are present in exfoliated samples and the different structures are associated with distinctive lineshapes in the Raman 2D mode. The rhombohedral samples show a more asymmetric 2D feature with an enhanced peak and shoulder, compared with the feature seen in Bernal samples. Taking advantage of this difference in lineshape, we were able to visualize stacking domains in exfoliated tri- and tetra-layer graphene with submicron spatial resolution. Even in samples of completely homogeneous layer thickness, we observed domains of different stacking order, with approximately 15% of the total area displaying rhombohedral stacking.

In our experiment, we prepared FLG samples by mechanical exfoliation of kish graphite (Toshiba) on both bulk $SiO_2$ (Chemglass, Inc) and Si substrates covered with a 300-nm-thick oxide layer. The substrates were cleaned by etching in piranha (sulfuric acid and hydrogen peroxide) solution. The typical area of our graphene samples varied from several hundreds to thousands of $\mu m^2$. We first examined the samples by IR spectroscopy. This technique permits accurate determination of layer thickness in FLG and, through the differences in the low-energy electronic structure, also of the stacking order[4]. We observed two distinct groups of IR spectra both for the trilayer and tetralayer graphene, corresponding to Bernal (ABA) and rhombohedral (ABC) stacking (see supplementary material and the reference[4] for details). We then performed Raman measurement on the same FLG samples. Raman spectra were collected in a backscattering geometry using linearly polarized laser radiation at wavelengths (photon energies) of 633 nm (1.96 eV), 597 nm (2.09 eV), 514 nm (2.41 eV), and 458 nm (2.71 eV). The laser beam was focused to a spot size of ~1 μm on the graphene samples. We obtained Raman spatial maps for an excitation wavelength of 514 nm by raster scanning with a precision two-dimensional stage having a step size of 0.5 μm or 1 μm[33]. For such spatial mapping of the Raman response, we generally used a spectral resolution of ~8 $cm^{-1}$ (obtained with a 600 grooves/mm grating). For the measurement of key spectra, however, a spectral resolution of ~2 $cm^{-1}$ (1800 grooves/mm grating) was chosen to elucidate the details of the lineshape.

We observed consistent differences in the lineshape of the Raman 2D-mode between samples with Bernal and rhombohedral stacking. Figure 2 displays results for trilayer samples. For all excitation photon energies, the ABC trilayers displayed more asymmetric and broader lines than ABA trilayers. In particular, we observed a sharp peak and an enhanced shoulder in the ABC spectra for all excitation photon energies (More detailed analysis of the spectra can be found in the supplementary material). This signature of stacking order is clear in all pristine trilayer samples. (We note that chemically processed samples may exhibit broadened 2D-mode spectra from doping and disorder. This may obscure the stacking-order signature.)

While the 2D mode has been applied widely to identify mono- and bi-layer samples[6, 25, 31, 32], its application to analysis of trilayers has been more limited because of the lack of consensus on the 2D lineshape of different trilayer samples. This mode was found to have a more asymmetric shape in some studies[6, 34, 35] and a less asymmetric shape in others[32]. Our results suggest that such variation may arise from the difference in stacking order among samples.

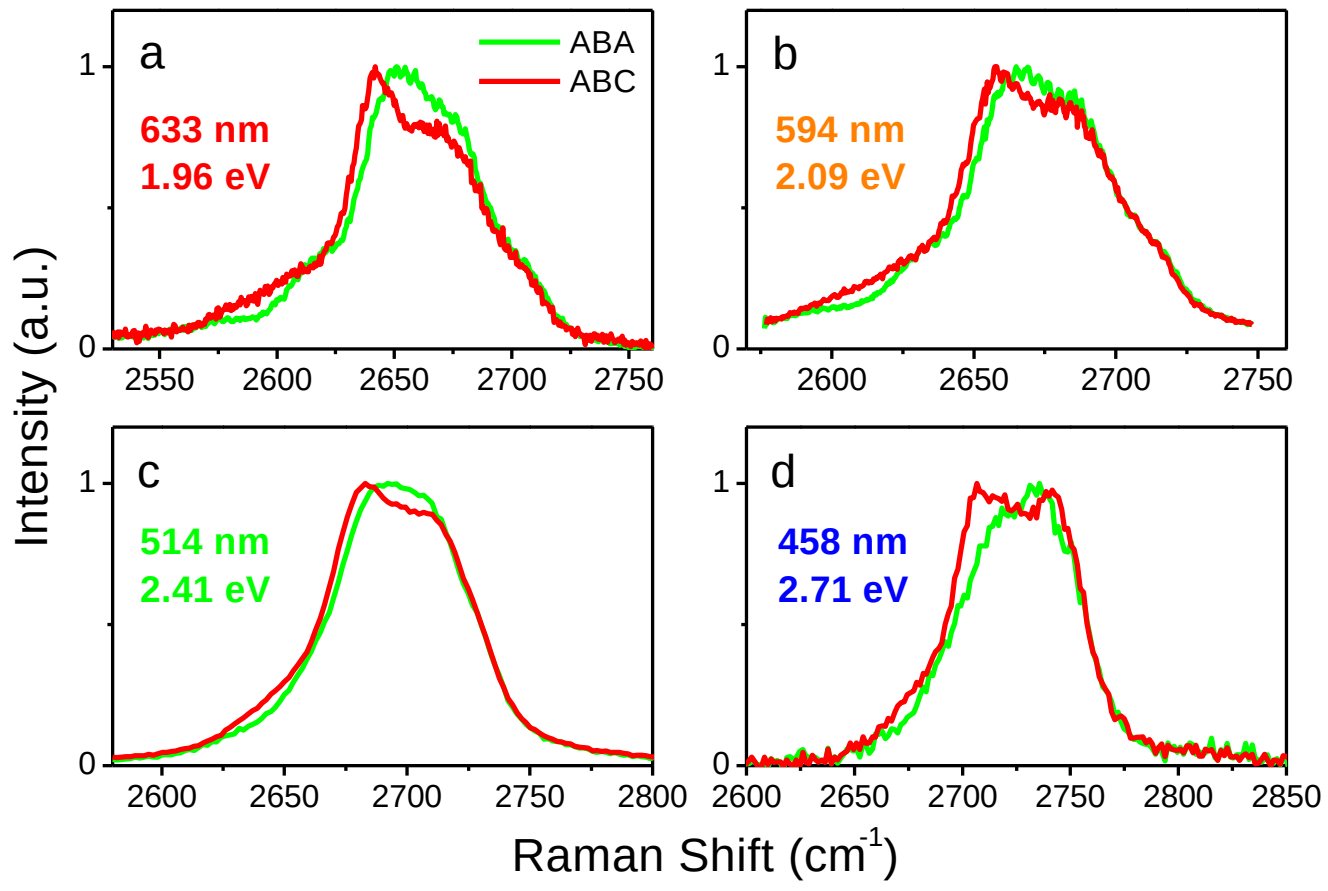

**Figure 2.** Raman spectra of the 2D-mode of ABA (green line) and ABC (red line) trilayer graphene samples at four different laser excitation wavelengths. The increase in the average Raman shift with excitation photon energy is an expected consequence of the double-resonance process that selectively couples phonons with different momenta around the *K*-point.

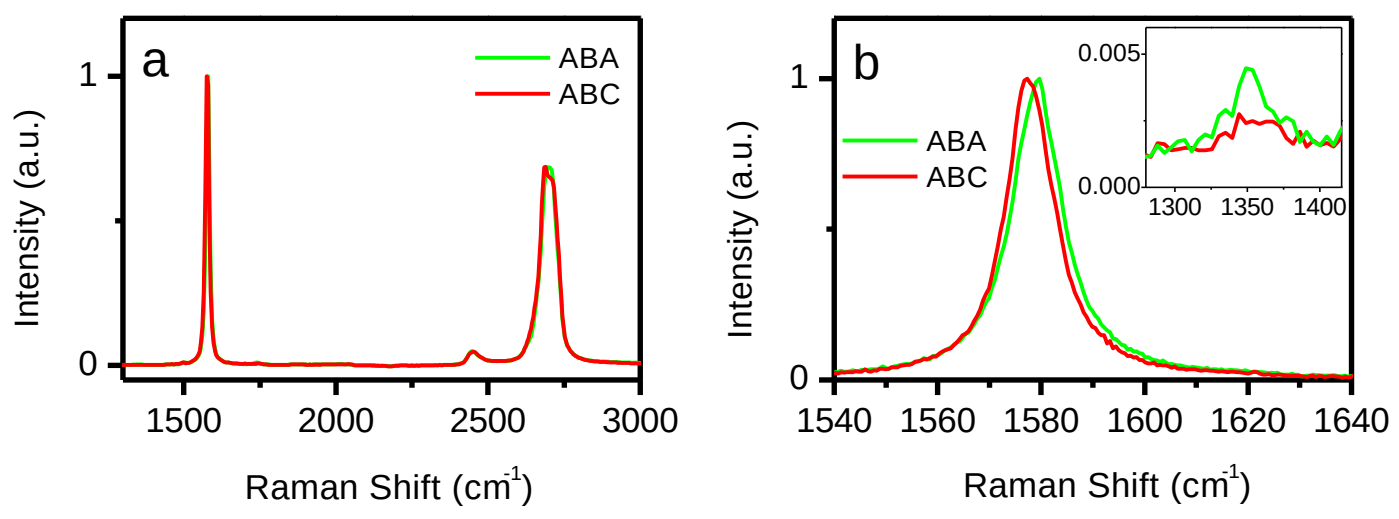

**Figure 3.** Raman spectra of graphene trilayers with ABA (green line) and ABC (red line) stacking order. (a) Raman spectra over a broad energy range. (b) Details of G-mode and D-mode (inset) spectra. The excitation laser wavelength is 514 nm.

We have also examined the Raman G-mode of ABA and ABC trilayer graphene (Figure 3). The spectra were taken in ABA and ABC stacking domains of a single trilayer sample to maintain similar doping and strain condition. (Detailed information about the coexistence of such stacking domains is discussed below.) Only slight differences are observable in the G-mode frequency and lineshape, as well as in the 2D/G ratio. As G-mode is not influenced by electronic resonances, we ascribe the small red shift (~1 cm$^{-1}$) of the G-mode frequency of ABC trilayer compared to ABA trilayer to the slight difference of their phonon band structures[36]. In addition, the very weak D-mode feature [D/G<0.005, as shown in the inset of Figure 3b] indicates the high crystalline quality in the trilayer areas of either stacking order.

The similarity of the G-mode features in ABA and ABC trilayers implies that the greater sensitivity of the 2D mode to stacking order is the result of differences in electronic structure. The Raman 2D mode is expected to be affected by the electronic properties since it arises from a double-resonance process that involves transitions among various electronic states[6, 25, 29, 30]. It is this sensitivity that has rendered the 2D-mode a fingerprint for mono- and bi-layer graphene samples[25, 31].

In order to visualize the spatial distribution of the stacking domains, we implemented a method for Raman mapping. To this end, we needed to define a quantity that could effectively encode the differences between the Raman spectra for the two different stacking orders. We examined several schemes, including ones based on changes in the centroid and asymmetry of the Raman spectrum. We found, however, that using the spectral width of the 2D mode captured the differences in a simple and robust fashion. To extract the width, we fit the spectrum at each pixel in the spatial mapping to a single Lorentzian function. We then produced spatial images by displaying the full width at half maximum (FWHM) of the fit function for each pixel in the image. A direct determination of the width from the spectra can also be used to generate spatial maps. This procedure led, however, to noisier images.

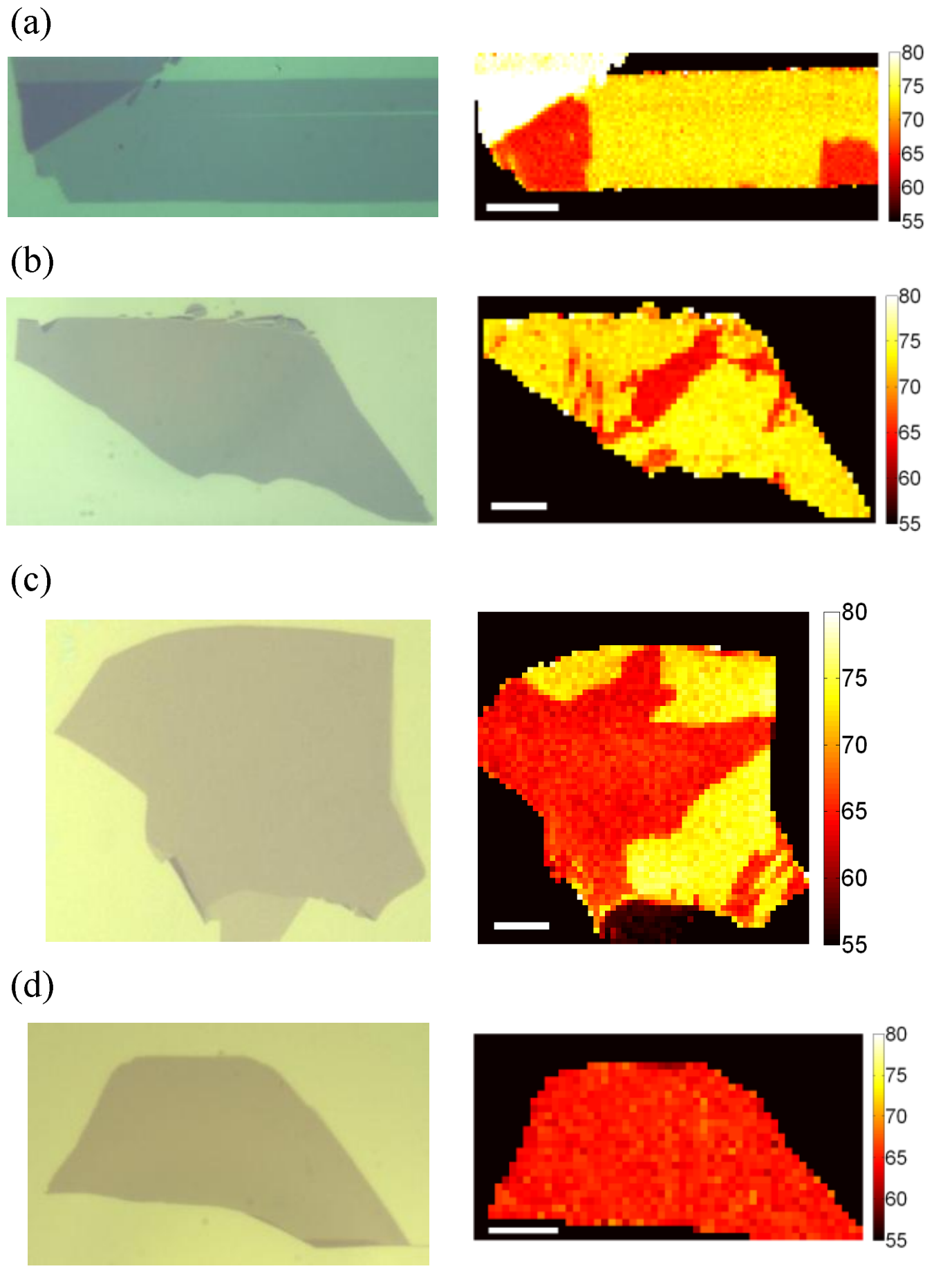

**Figure 4.** Optical images and spatial maps of the spectral width of Raman 2D-mode feature for trilayer graphene samples. The homogeneity of the optical images shows the uniformity of the layer thickness of the four samples. The Raman images, taken with 514-nm excitation, are color coded according to the width of the Raman feature (FWHM in units of cm$^{-1}$). The red and yellow regions in the images correspond, respectively, to ABA and ABC trilayer graphene domains. The step size in the scans is 0.5 μm for (a), and 1 μm for (b)-(d). The scale bars are 10 μm in length. Similar results can be obtained using other wavelengths for the excitation laser.

Figure 4 presents examples of Raman mapping of the trilayer graphene samples that exhibit domains of differing stacking order. For comparison, we also show optical images of these samples. No difference in the optical contrast is observed across the full area of the samples, indicating that each sample is entirely homogeneous in thickness. We have further performed IR measurements to confirm the results of the Raman mapping. The IR spectra obtained in the different regions of these four samples corresponds precisely to those of ABA and ABC trilayers (Figure S1 in the supplementary material).

We note that exfoliated graphene samples exhibit spatially inhomogeneous carrier doping effects when deposited on typical insulating substrates[33, 37]. Could this charging effect influence our Raman data? To address this issue, we investigated a trilayer sample with ABC stacking that was suspended over a quartz trench and, hence, isolated from any substrate effects. There was no observable change in the Raman 2D lineshape between the free-standing sample that has negligible doping and samples supported on the quartz substrates that have estimated unintentional doping of $n \sim 5 \times 10^{12}$ $cm^{-2}$ (see supplementary material). We therefore conclude that the lineshape of the 2D-mode is not sensitive to a variation of doping density in the order of $10^{12}$ $cm^{-2}$.

The coexistence of ABA and ABC stacking order is striking. Among the 45 trilayer samples that we prepared, 26 exhibited purely ABA stacking, while 19 displayed mixed ABA and ABC stacking. None of the samples showed purely ABC stacking. In the 19 samples of mixed stacking order, only 5 samples contain large (>200 $\mu m^2$), homogeneous regions of ABC-stacking order, as in Figure 4a, c. If we consider the total area associated with the two different stacking orders, we find that ~ 85% area of our samples corresponds to ABA stacking and ~ 15% to ABC stacking. This result is comparable to that obtained in earlier x-ray diffraction studies of graphite[38, 39], which indicate that graphite typically contains 80% of the Bernal structure, 14% of the rhombohedral structure, and 6% of a disordered structure[38]. The similarity of our results suggests that the different stacking orders observed in trilayer graphene originate from the pristine structure of the graphite used in the exfoliation process, which is not modified during the exfoliation of the layers. This claim is supported by the complicated patterns of stacking domains in our samples. One would not expect these patterns to be produced by mechanical processing.

The method of imaging stacking order by Raman spectroscopy can be generalized to investigations of FLG of greater thickness. Here we show the results for tetralayer graphene. Figure 5 shows the 2D-mode Raman spectra for ABAB (green line) and ABCA (red line) tetralayer graphene (Detailed information on the stability and IR spectroscopy of these two types of tetralayer graphene can be found in the references[4, 10]). By carefully examining the spectra, we observed distinct lineshapes for the two stacking orders. ABCA tetralayers show more structured, asymmetric lines with greater widths than the ABAB tetralayers. In particular, we observed a sharp peak and an enhanced shoulder in the ABCA spectrum at 2680 $cm^{-1}$ and 2640 $cm^{-1}$. Such distinctions in the 2D-mode spectra are observed in all ABAB and ABCA tetralayer samples. We note that both the ABC trilayer (Figure 2c) and the ABCA tetralayer exhibit similar 2D-mode lineshapes, indicating that the asymmetric and broadened features are characteristics of this stacking order.

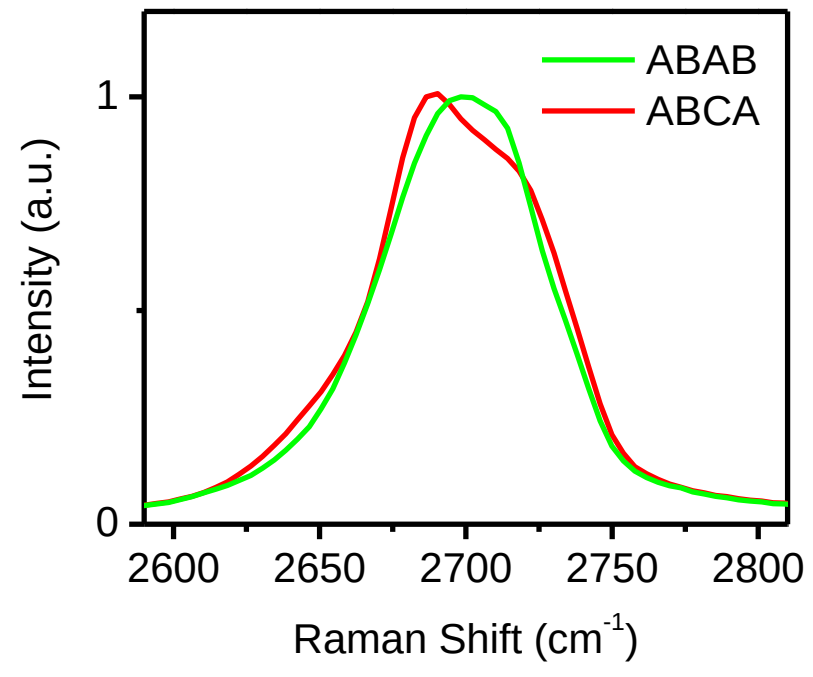

**Figure 5.** Raman 2D-mode spectra for the tetralayer graphene samples of ABAB (green line) and ABCA (red line) stacking order.

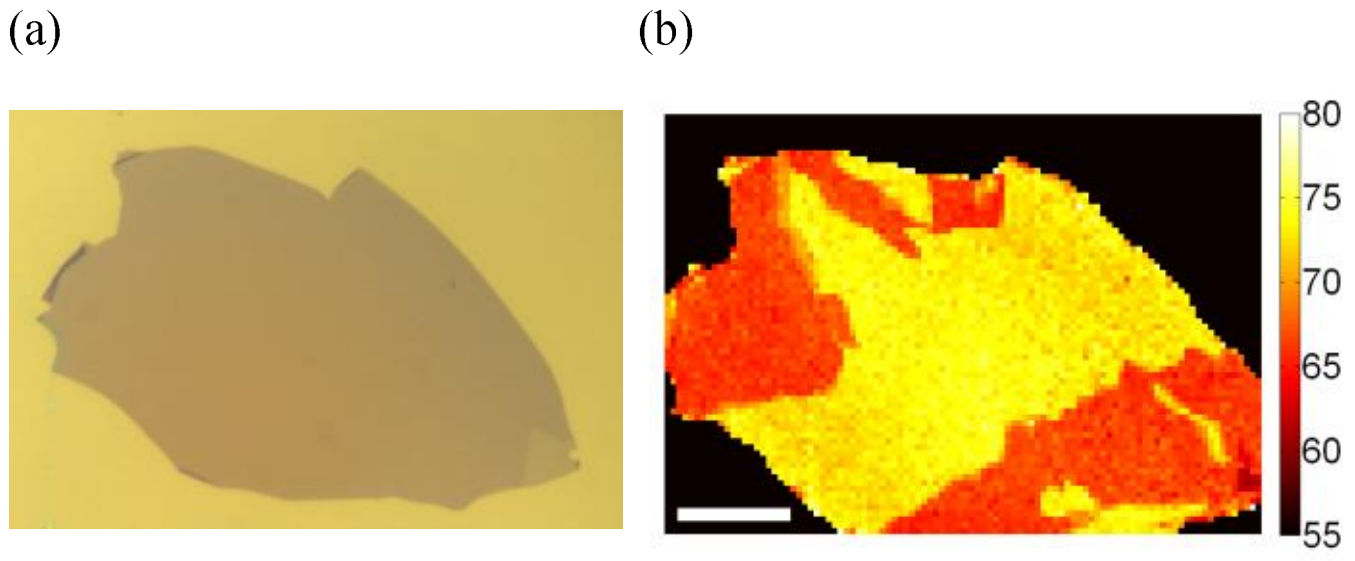

**Figure 6.** (a) Optical and (b) Raman images of a specific tetralayer graphene sample. The optical image shows the uniformity of the layer thickness of the sample. The Raman map of the spectral width of Raman 2D-mode exhibits domains with different stacking order. The Raman images, obtained with 514-nm excitation, are color coded according to the FWHM of the Raman feature (in units of $cm^{-1}$). The red and yellow regions in the images correspond, respectively, to tetralayer graphene with ABAB and ABCA stacking. The step size for the Raman mapping was 1 μm. The length of the scale bar is 20 μm.

Fig. 6 (a) and (b) show, respectively, the optical image and Raman image of a tetralayer graphene sample with mixed (ABAB and ABCA) stacking order. The homogeneous optical image indicates that the sample is entirely graphene of 4 layers in thickness. The Raman image, which encodes the 2D-mode spectral width (FWHM) extracted from a single Lorentzian fit, shows sharp contrast for regions of different stacking order. The stacking sequences determined by Raman spectroscopy in these domains were further confirmed by IR spectroscopy.

The characteristics of the stacking domains based on an analysis of 56 samples, were found to be similar to those for trilayer graphene. In particular, the ratio of the area of ABAB to ABCA stacking was 85:15. The similarity of the domains in tetralayer and trilayer graphene confirms the common origin of the different stacking sequences, namely, the stacking order of the kish graphite, which remains unchanged during the exfoliation process.

The ability to directly visualize the domains of stacking order provides a means of assessing the thermodynamic stability of the different structures. Our Raman images show that Bernal and rhombohedral stacking order can coexist in micron-sized domains of trilayer thickness at room temperature. By annealing the samples in an argon environment, we found, by both Raman and IR spectroscopy, that the domains of rhombohedral stacking order are stable up to 800 ºC, the maximum annealing temperature in our experiment. This result is consistent with the stability of bulk rhombohedral graphite to over 1000 ºC[40, 41]. Our result shows that this stability is retained even for *atomically thin* rhombohedrally stacked crystallites (More detailed information on the annealing measurements can be found in the supplementary material).

In conclusion, we have demonstrated Raman spectroscopy to be an effective tool for the characterization of stacking order in few-layer graphene. Bernal (ABA) and rhombohedral (ABC) tri- and tetra-layer graphene samples exhibit clear differences in the lineshape and width of the Raman 2D line. By Raman spatial mapping, we find that for typical exfoliated tri- and tetra-layer samples about 15% of the area has rhombohedral stacking rather than the usual Bernal stacking. The domains of rhombohedral stacking are generally only of micron length. The Raman technique presented in this paper should accelerate the research on FLG. For instance, various studies[10, 12, 15, 19] have predicted that a significant and electrically tunable band gap can be opened in rhombohedral trilayer graphene by the application of an electric field. However, for Bernal trilayers with their differing symmetry, this effect will be much smaller. With the Raman technique presented here, we can readily identify the domains of stacking in FLG and can test these predictions experimentally by constructing an appropriate gated device. In addition to assuring the desired crystal structure of samples, the Raman mapping capabilities allow identification of domain boundaries. This information should permit detailed exploration of the structural and electronic properties at these interfaces.

**Acknowledgements** We thank K. F. Mak, L. M. Malard, S. Ryu and J. Shan for discussions. We also acknowledge support from the Office of Naval Research under the MURI program, from DOE Basic Energy Sciences under grants DE-FG02-98ER14861 and DE-FG02-07ER15842, from the National Science Foundation under grant CHE-0117752 and the NRI program of the SRC, and from the New York State Office of Science, Technology, and Academic Research (NYSTAR).

**Supporting Information Available.** The material contains the results of the IR spectroscopy, detailed analysis of the Raman spectra of ABA and ABC graphene trilayers, information of the influence of the substrate on the 2D-mode lineshape, and additional discussion about the thermal stability of ABC stacking order.

# Supplementary Material of
# "Imaging Stacking-Order in Few-Layer Graphene"

## 1. Infrared Measurement

In our experiment, the exfoliated few-layer graphene (FLG) samples on bulk $SiO_2$ (quartz) and Si/$SiO_2$ substrates were first examined by infrared (IR) spectroscopy. The measurements were performed in both reflection and transmission geometry using a micro-Fourier Transform Infrared spectrometer with a globar source and a HgCdTe detector. As the intrinsic infrared response of FLG on bulk $SiO_2$ and Si/$SiO_2$ substrates is the same, here we only present results for the bulk $SiO_2$ substrate since the analysis of the optical measurements is simpler. To determine the optical sheet conductivity σ(ħω) of the FLG samples as a function of photon energy $\hbar\omega$, we follow the same method as our previous work in monolayer [S1] and tetralayer [S2] graphene. We recorded the reflectance spectra of both the FLG films on the quartz substrate ($R_{FLG}$) and of the bare substrate ($R_{sub}$). We obtain the optical conductivity $\sigma(\hbar\omega)$ directly from the fractional change of the reflectance as

$$\delta_R = \frac{R_{FLG} - R_{sub}}{R_{sub}} = \frac{4}{n_{sub}^2 - 1}\frac{4\pi}{c}\sigma .$$

Here $c$ denotes the speed of light in vacuum and $n_{sub}$ is the frequency-dependent refractive index of the quartz substrate.

The IR optical conductivity provides an effective probe to the electronic structure of FLG. While the low-energy (< 0.7 eV) conductivity reflects the details of electronic structure and doping level, the high-energy (> 0.7 eV) part provides a precise identification of layer thickness. For photon energies well above the interlayer coupling (~0.4 eV), FLG graphene behaves much like independent graphene monolayers and its optical conductivity is nearly independent of the stacking sequence. Since graphene monolayer has an optical conductivity of $\pi e^2/2h$ in this spectral range, we can identify the trilayer graphene by the expected conductivity value of $3\times\pi e^2/2h$. Taking this as guidance, we found in total 45 trilayer graphene samples.

We observed two distinctive groups of IR response in the optical conductivity spectra of trilayer graphene, as shown in Figure S1 (a). The first kind of spectrum (green line) shows an absorption peak at 0.53 eV, which matches the result of ABA trilayer graphene [S3]. The second kind of spectrum (red line) exhibits two narrow peaks at 0.33 and 0.39 eV. This distinct IR response spectrum of trilayer graphene, which is found in ~ 10% of our samples, has not been reported previously. It indicates the presence of a low-energy electronic structure different from that of ABA trilayer graphene. In addition, we have also observed IR spectra with both the features of the previous two kinds of spectra. This third type of spectrum, found in ~30% of our trilayer samples, can be described as a linear combination of the first two kinds of spectra [Figure S1 (b)].

These observations immediately lead to the consideration of the different crystallographic stacking sequences in trilayer graphene. The possible low-energy arrangements of adjacent layers of graphene are obtained by displacement of one layer along the direction of the honeycomb lattice by a carbon-carbon bond length. We associate these two basic types of spectra with the two distinct low-energy crystallographic structures of trilayer graphene [S4]: ABA (Bernal) or ABC (rhombohedral) stacking (Figure 1 in the paper). The existence of different stacking order in FLG and its strong impact on electronic structure have been demonstrated experimentally by IR spectroscopy [S2]. Following this work we assign the first and second kind of spectrum to trilayer graphene with ABA and ABC stacking order. The third kind of spectrum is attributed to the trilayer samples with mixed stacking order.

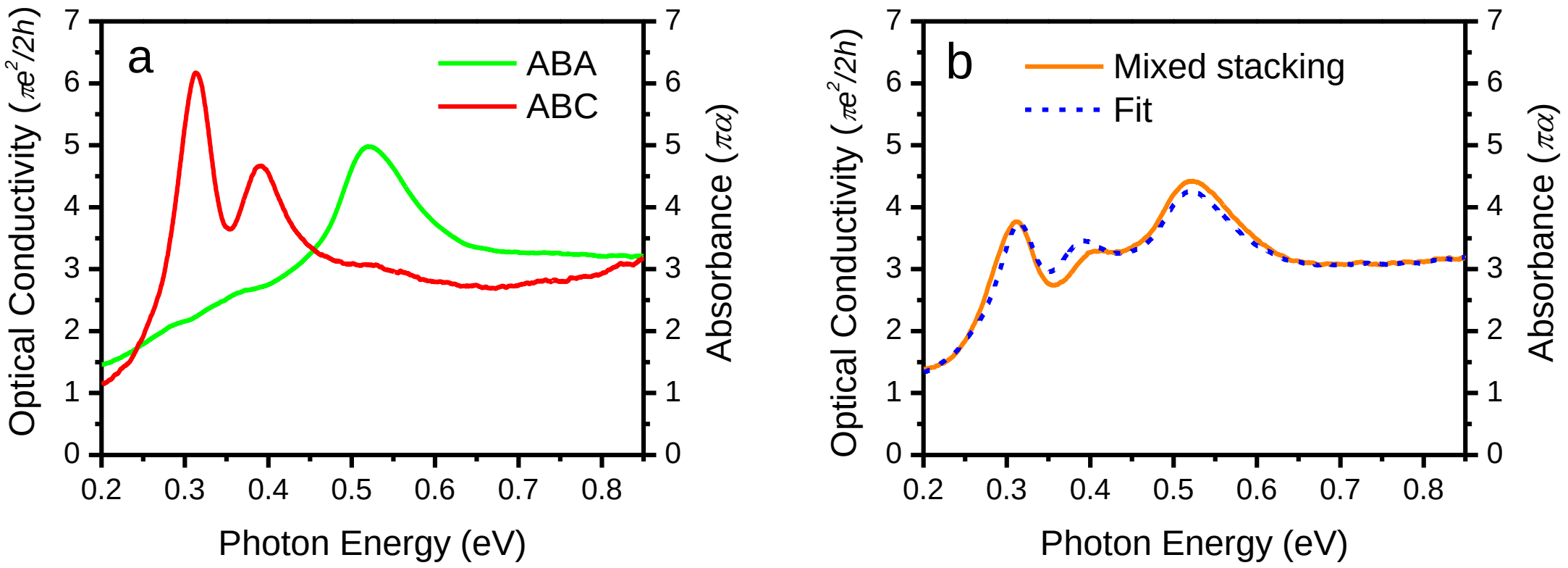

**Figure S1.** Optical conductivity of different trilayer graphene samples. (a) Spectra of trilayer with ABA (green line) and ABC (red line) stacking order. (b) The spectrum of a trilayer sample with mixed stacking order (orange line). The spectrum can be described as a linear combination of 67% ABA stacking and 33% ABC stacking (dashed blue line). The slight discrepancy may reflect different doping and strain levels.

## 2. Analysis of trilayer 2D-mode Raman spectra

As shown in Figure 2 in the paper, ABA and ABC trilayer graphene exhibit distinct 2D-mode Raman lineshapes. The 2D band arises from a double-resonance process that involves inter-valley (i.e. between the K and K' point) scattering in the Brillouin zone and resonant electronic transitions. As trilayer graphene has three valence and three conduction bands, many electronic transitions can contribute to the 2D band. A recent study by group theory shows that up to 15 peaks in the 2D band are possible in ABA trilayer graphene [S5]. In practice, however, we may consider fewer transitions, since many of them have close energy separations.

We have fit the 2D-mode Raman spectra of ABA and ABC trilayer graphene with multiple Lorentzian functions. The FWHM of all the Lorenztian functions are fixed to be the same as that of the 2D band of monolayer (~25$cm^{-1}$). We only vary the peak positions and intensities. We found that a good fit can be achieved with 6 Lorentzian functions. Figure S2 (a-h) shows such for both ABA and ABC trilayer graphene samples for all excitation energies. The differences in the spectra and, correspondingly, in the fitting parameters, become more prominent as the excitation photon energy decreases. This trend presumably reflects the more pronounced differences in electronic structure at low energies for the two stacking orders.

We have also considered the shift of the 2D band as a function of the excitation energy. We extract the mean Raman shift by averaging the 2D band spectra weighted with the spectral intensity. As shown in Figure S3, ABA trilayer graphene has a higher mean 2D-band Raman shift than ABC trilayer, but a very similar dispersion.

To investigate substrate effects on our method, we performed the Raman measurements on free-standing ABC-stacked trilayer graphene samples. The samples are prepared by mechanical exfoliation of kish graphite on quartz substrates with pre-patterned trenches. Some parts of the ABC trilayer graphene covering the trenches are suspended. They are thus isolated from any perturbation induced by the substrates. Figure S4 shows the Raman 2D-mode spectra recorded for supported and suspended parts of a single ABC trilayer sample. Apart from a slight shift of frequency, both spectra show essentially the same lineshape. The difference of FWHM obtained by single-Lorentzian fits is within 1$cm^{-1}$. According to the literature [S6-9], such substrates typically induce an unintentional doping of $n$ ~ 5 x $10^{12}$ $cm^{-2}$ in graphene, which varies from sample to sample and is also inhomogeneously distributed on a submicrometer scale within a given graphene sample. Our results show that the corresponding changes in the 2D-mode lineshape are very slight and do not impair our ability to distinguish between ABA and ABC stacking order.

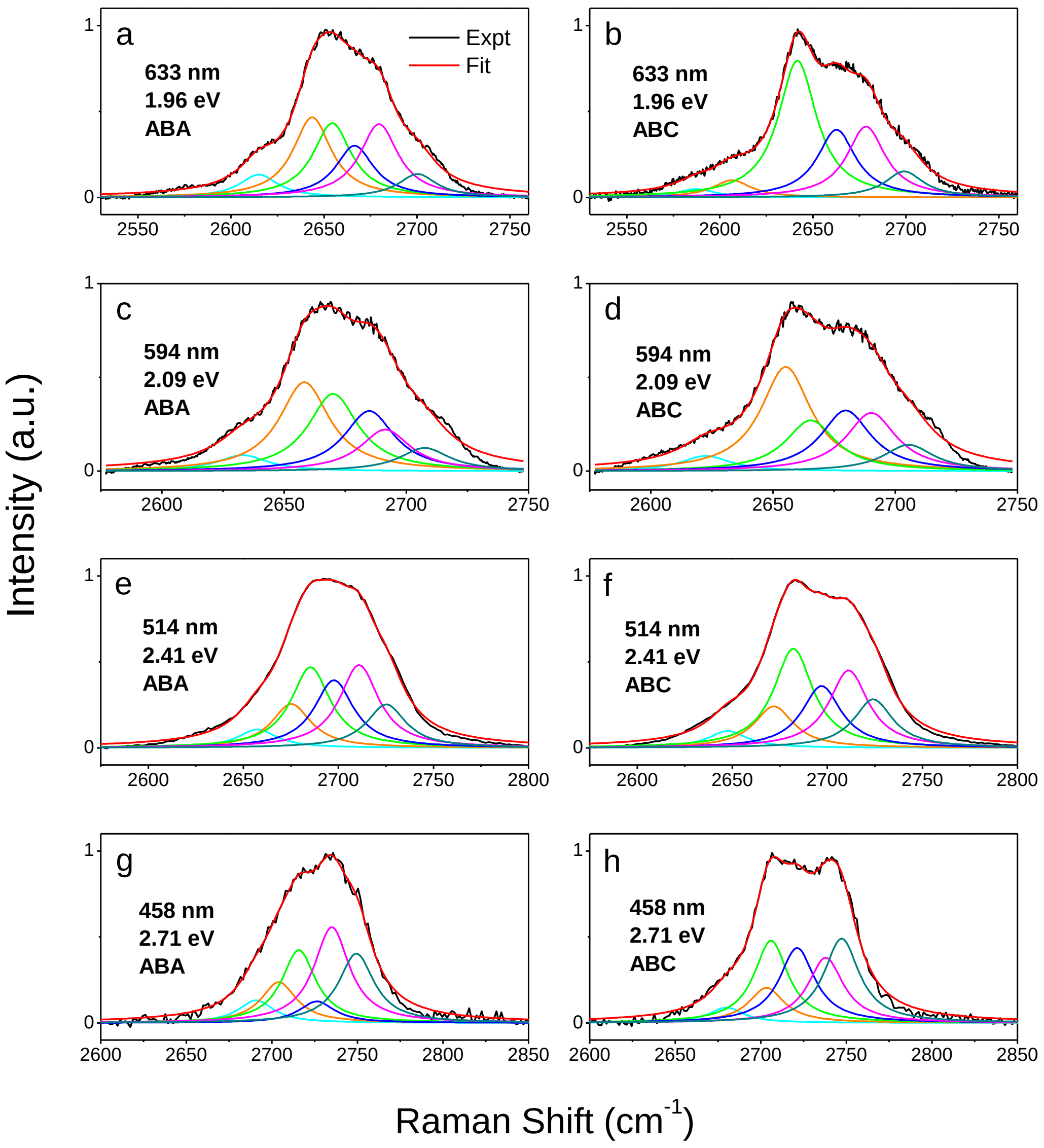

**Figure S2.** 2D-mode Raman spectra of graphene trilayers with ABA (left) and ABC (right) stacking order at different excitation energies. The black lines are experimental data. The red lines are fits by 6 Lorentzian functions, all with a FHWM of 25 $cm^{-1}$. The lines of other colors are the Lorentzian components of the fits.

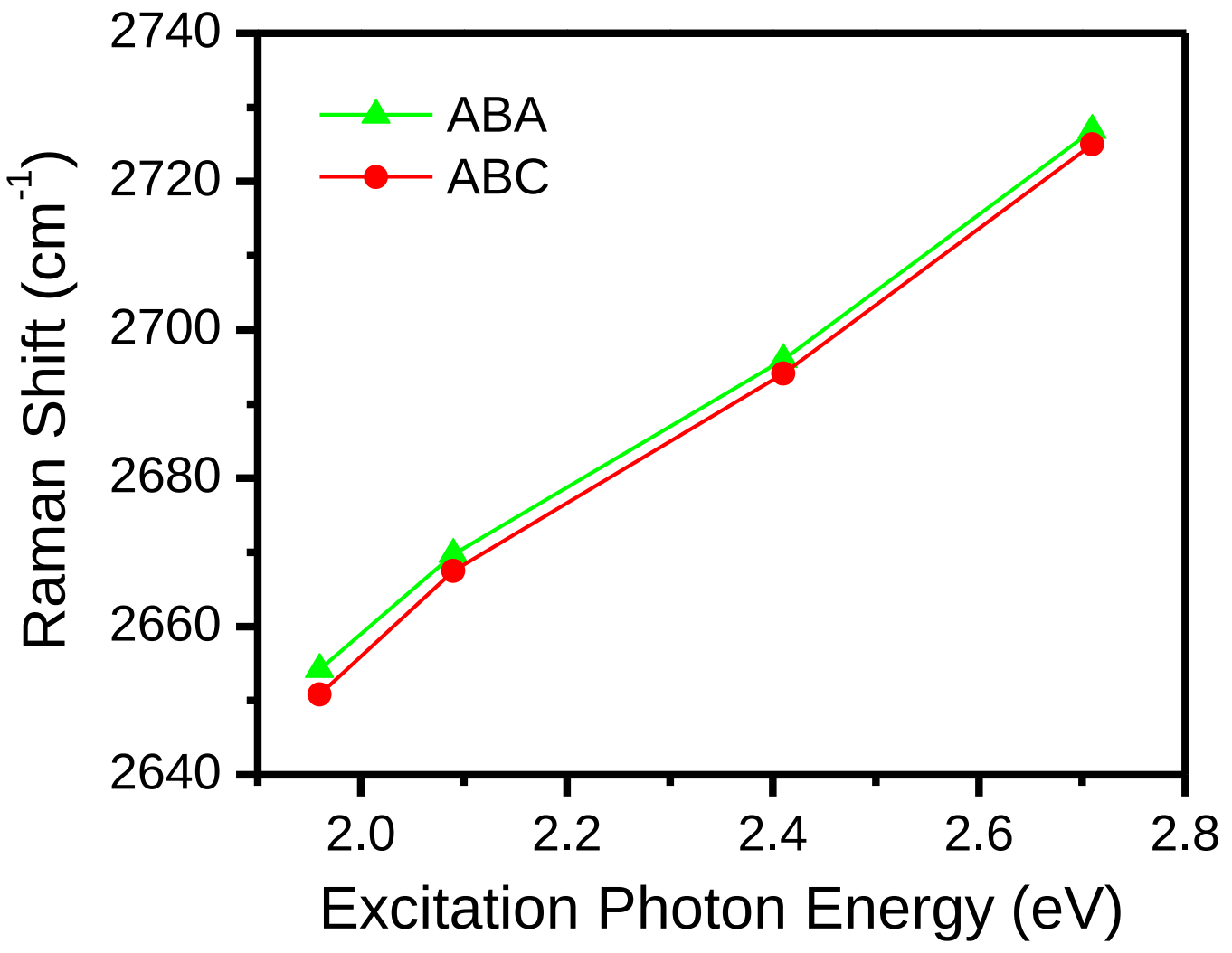

**Figure S3.** Mean Raman shift of the 2D-mode features for graphene trilayer with ABA (green triangles) and ABC (red dots) stacking order for different excitation laser energies. The data are obtained by averaging the Raman shifts weighted by the corresponding spectral intensity in the 2D-mode spectra.

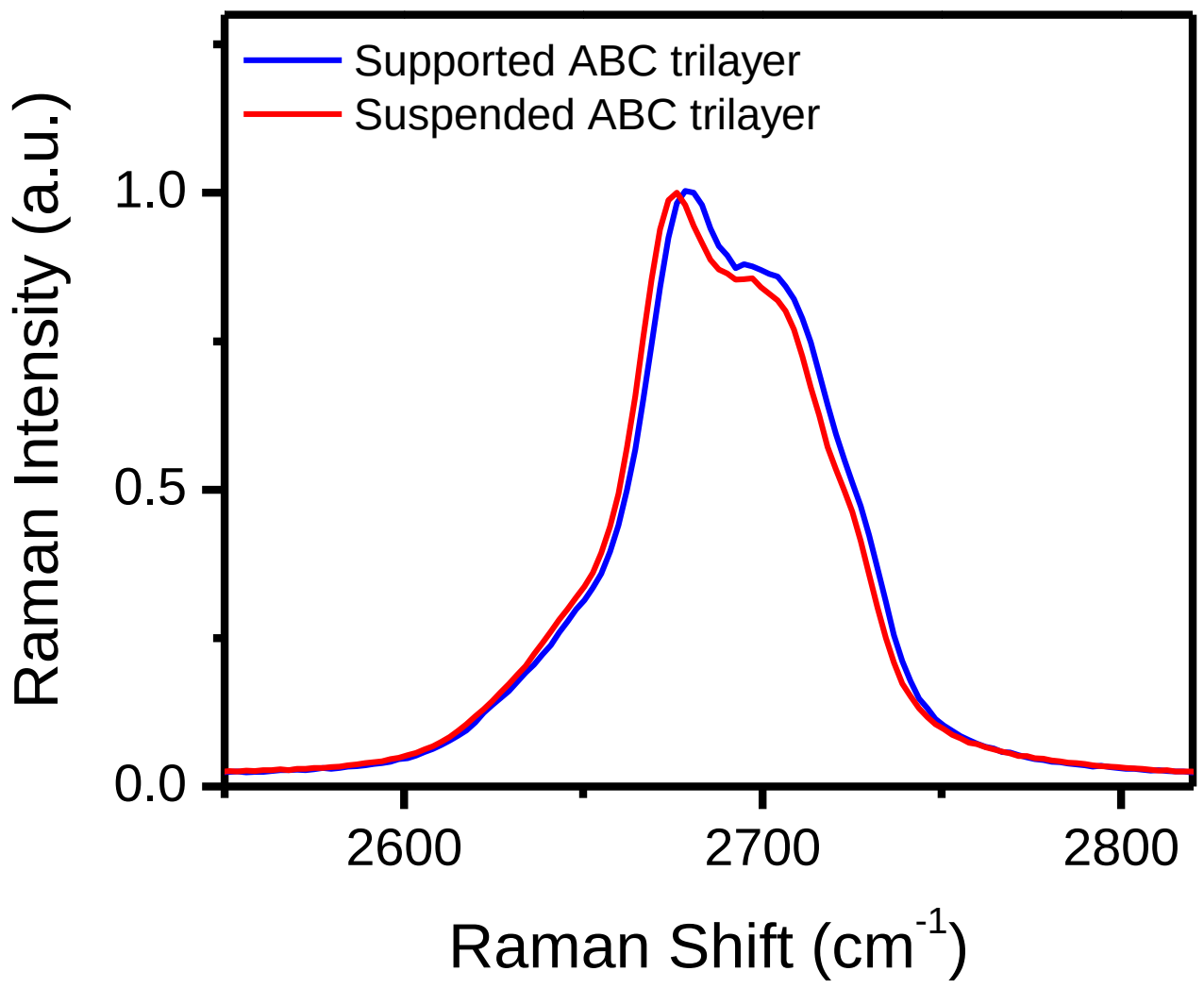

**Figure S4.** Raman 2D-mode spectra from an ABC trilayer sample supported on a quartz substrate (blue line) and suspended over a trench. The two spectra show very similar lineshapes and widths. The difference of FWHMs obtained by the single-Lorentzian fits is less than 1 $cm^{-1}$. The excitation laser wavelength is 532 nm.

## 3. Thermodynamic stability of ABC stacking order

We investigated the thermodynamic stability of the ABC stacking order in trilayer graphene by annealing the samples to high temperatures. We first chose a piece of pristine trilayer sample with mixed stacking order and examined its domains of stacking by the Raman mapping of the FWHM of the Raman 2D mode [Figure S5 (a), which is the same as Fig. 4(c) in the main body of the paper]. The ABA and ABC domains are clearly encoded as the red and yellow colored regions in the Raman image. We then annealed the same sample in an argon environment at different temperatures and subsequently examined the structure of the domains by Raman mapping. We found that the domains of ABC stacking remained stable up to 800 ℃, the maximum annealing temperature in our experiment. Figure S5(b) displays the Raman image after annealing the sample at 500 ℃ for 10 hours. The domains of ABA and ABC stacking, still recognizable as the red and yellow regions, are unaltered. We note that the image contrast in the Raman mapping between the domains of ABA and ABC stacking is reduced by the thermal processing cycle. We attribute this to the introduction of excess doping and disorder in the sample, leading to broadening of the Raman 2D feature. We further confirmed the stacking order by IR spectroscopy, as discussed above in Sect. 1 of the Supplementary Material.

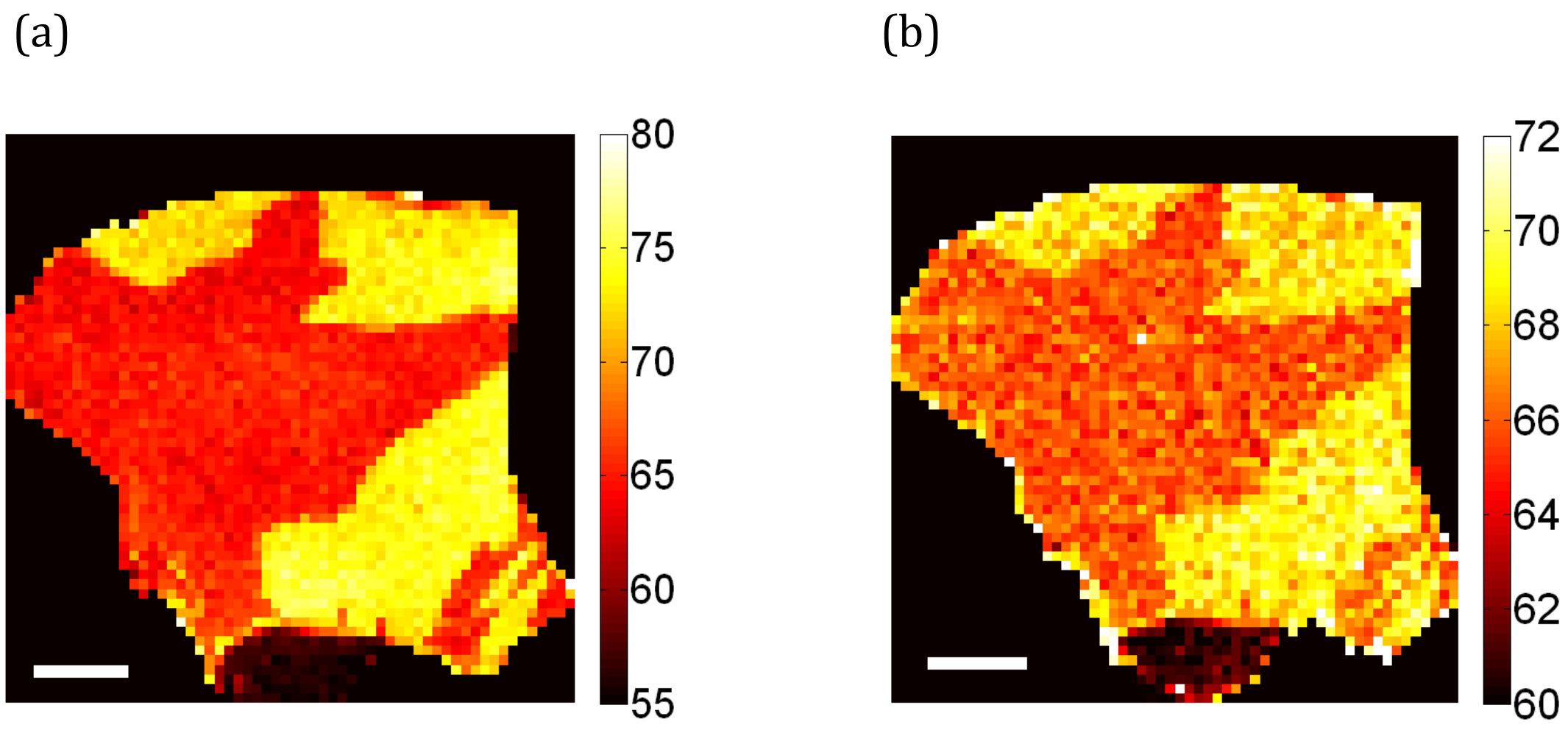

**Figure S5.** Influence of thermal annealing on the domains of different stacking order in trilayer graphene. Panels (a) and (b) display spatial maps of the spectral width of Raman 2D-mode feature for a pristine trilayer graphene sample (a) and for the same sample after thermal annealing in an argon environment at 500 ℃ for 10 hours (b). The Raman images are color coded according to the width of the Raman feature (FWHM in units of $cm^{-1}$). The red and yellow regions in the images correspond, respectively, to ABA and ABC trilayer graphene domains. The scale bars are 10 μm in length.